\documentclass[twocolumn,showpacs,nofootinbib,aps,prd]{revtex4}

\usepackage{graphicx,url,color}
\usepackage{amsmath,amsthm,amssymb}

\renewcommand{\vec}[1]{\boldsymbol{#1}}

\newcommand{\be}{\begin{equation}}
\newcommand{\ee}{\end{equation}}
\newcommand{\ba}{\begin{eqnarray}}
\newcommand{\ea}{\end{eqnarray}}

\def\lsim{\raise0.3ex\hbox{$\;<$\kern-0.75em\raise-1.1ex\hbox{$\sim\;$}}}
\def\gsim{\raise0.3ex\hbox{$\;>$\kern-0.75em\raise-1.1ex\hbox{$\sim\;$}}}

\def\theta{\vartheta}

\def\sigv{\langle \sigma v\rangle}
\begin{document}

\title{Antideuterons from dark matter annihilations and hadronization model dependence}

\author{L.~A.~Dal$^1$}
\author{M.~Kachelrie\ss$^{2}$}

\affiliation{$^1$Department of Physics, University of Oslo, Norway}
\affiliation{$^2$Department of Physics, NTNU, Trondheim, Norway}

\begin{abstract}
We calculate the antideuteron yield in dark matter annihilations on
an event-by-event basis using the HERWIG++ Monte Carlo generator.
We present the resulting antideuteron fluxes for quark and
gauge boson final states. 
As deuteron production in the coalescence model depends on momentum 
differences between nucleons that are small compared to 
$\Lambda_{\rm QCD}$, it is potentially very sensitive to the 
hadronization model employed.
We therefore compare our antideuteron yields to earlier results based on 
PYTHIA, thereby estimating their uncertainties.
We also briefly discuss the importance of $n>2$ final states for
annihilations of heavy DM particles.
\end{abstract}


\maketitle

%%%%%%%%%%%%%%%%%%%%%%%%%%%%%%%%%%%%%%%%%%%%%%%%%%%%%%%%%%%%%%%%%%%%%%%%%%%%%%
\section{Introduction}

Despite various cosmological and astrophysical indications for the presence 
of nonbaryonic dark matter (DM), its particle nature has yet to be proven. 
Restricting the space of candidates to 
thermal relics, the thermally averaged annihilation cross section
$\sigv$ at freeze-out is fixed by the DM abundance, 
while the DM mass $M_{\rm DM}$ is only weakly constrained: Unitarity of the 
$S$-matrix 
constrains the mass of any thermal relic as $m\lsim 50\,$TeV~\cite{GKH},
while the requirement that the DM is cold translates for  thermal relics 
into a lower mass limit of the order 10\,keV. 

One strategy towards DM detection is to search for its self-annihilation (or 
decay) products. The annihilation of symmetric DM leads to equal injection
rates of matter and antimatter particles into the Galaxy, while the 
cosmic ray flux 
from astrophysical sources is matter-dominated. A possible
way to detect DM is therefore to carefully estimate the expected
antimatter fluxes from astrophysical sources, and then to search for any
excess. The authors of Ref.~\cite{Donato:1999gy} suggested antideuterons as a signature
in addition to the usually discussed antiproton and positron signal
from DM annihilations. In particular, they argued that the DM antideuteron 
spectra are much flatter at low energies than those from cosmic-ray--gas
interactions.

The correct application of the coalescence model in deuteron production
requires its implementation on an event-by-event 
basis~\cite{Kadastik:2009ts,master}. Since deuteron 
production depends on  momentum differences $p_0$ that are small or
comparable to $\Lambda_{\rm QCD}\sim 200$\,MeV, one may expect a rather strong
dependence of the results on the hadronization models employed
in the Monte Carlo simulation. Moreover, the coalescence model probes
not only the energy spectrum of nucleons, but also their two-particle correlations
in momentum space. Since the physics underlying different
hadronization models---{\it e.g.} cluster hadronization versus string 
fragmentation---varies strongly, the choice of hadronization model could 
 thus have a profound effect on the generated antideuteron spectra.

The main aim of this work is therefore to derive results for the
antideuteron yields calculated using HERWIG++~\cite{Gieseke:2011na} version 2.4.2
(based on a cluster hadronization model) and to compare them to earlier 
results using PYTHIA~\cite{Sjostrand:2007gs} (based on
string fragmentation).  We also briefly discuss
the importance of $n>2$ final states for
annihilations of heavy DM particles.

%%%%%%%%%%%%%%%%%%%%%%%%%%%%%%%%%%%%%%%%%%%%%%%%%%%%%%%%%%%%%%%%%%%%%%%%%%%%%%
\section{Deuteron production and hadronization}

The fusion of (anti)protons and (anti)neutrons into (anti)deuterons
is usually described with the so-called coalescence model. This model is 
based on the assumption that any nucleons with a momentum difference 
$\Delta p < p_{0}$ for some given $p_0$ will merge and form a nucleus. 
In its initial form, the model was applied to 
deuteron production in  nucleus-nucleus interactions. In the simplest
approximation, the final-state hadrons in nucleus-nucleus scattering
are formed in a ``fireball'' and are emitted close to isotropically
in the center-of-mass (CoM) frame. Assuming additionally that correlations
between nucleons are negligible, the antideuteron energy spectra in the 
lab frame can be derived from the 
energy spectra of nucleons as~\cite{old,Ka80}
\ba  \label{Coal} 
 \frac{dN_{\bar{d}}}{dT_{\bar{d}}}
 & = &
 \frac{p_0^3}{6} \frac{m_{\bar{d}}}{m_{\bar{n}} m_{\bar{p}}} 
 \frac{1}{\sqrt{T_{\bar{d}}^{2} + 2 m_{\bar{d}} T_{\bar{d}}}}  
 \frac{dN_{\bar{n}}}{dT_{\bar{n}}} \frac{dN_{\bar{p}}}{dT_{\bar{p}}} \,.
\ea
Here, $p_0$ is the maximal momentum difference allowed between a 
$\bar p\bar n$ pair in order
to form an antideuteron according the coalescence model, 
$m_d, m_p$, and $m_n$ denote the deuteron, proton and neutron masses 
respectively, 
and the nucleon spectra on the RHS are to be evaluated at the value 
$T_{\bar{p}} =T_{\bar{n}} = T_{\bar{d}}/2$ of the kinetic energy $T_i=E_i-m_i$.

The constant prefactor in Eq.~\eqref{Coal} varies depending on the
assumptions made in its derivation~\cite{master}. Such a factor can, 
however, be absorbed by re-defining $p_0$. For later reference, we note
that our definition of $p_0$ agrees with the one of Ref.~\cite{Kadastik:2009ts},
while the one of Refs.~\cite{Donato:2008yx,Brauninger:2009pe} differs by a
factor two.

The coalescence model can easily be implemented directly in a Monte Carlo 
simulation by comparing the momenta of the final state nucleons in their
respective CoM frames for each individual annihilation event.  We will from now 
on refer to results from calculations where 
coalescence was applied on an event-by-event basis within the Monte Carlo 
as ``Monte Carlo'' results, while results from calculations where 
coalescence was applied to the average antiproton and antineutron energy 
spectra  using Eq.~\eqref{Coal} will be referred to as ``isotropic'' results.  
Correspondingly, we refer to the two approaches as the Monte Carlo approach 
and the isotropic approach.

%%%%%%%%%%%%%%%%%%%%%%%%%%%%%%%%%%%%%%%%%%%%%%%%%%%%%%%%%%%%%%%%%%%%%%%%%%%%
\subsection{Determining $p_0$}

The momentum threshold $p_0$ for the formation of deuterons is found
in both approaches by running simulations and adjusting $p_0$ such that 
the computational result matches the experimental one.  
Antideuteron production in $e^+e^-$-collisions 
was studied by the ALEPH collaboration~\cite{Schael:2006fd} using LEP-I data.
At the $Z$-resonance, each hadronic event was found to give rise to 
$(5.9 \pm 1.8 \pm 0.5)\times 10^{-6}$ antideuterons in the momentum range 
$0.62\,{\rm GeV} < k_{\bar{d}} < 1.03$\,GeV in the angular 
acceptance range $|\cos\theta| < 0.95$ of the detector.

Our simulations using HERWIG++ reproduce the experimentally measured
antideuteron yield choosing $p_0=110$\,MeV in the ``Monte Carlo approach''
and $p_0=126$\,MeV in the ``isotropic approach'', respectively. Note that 
the $p_0$ values differ in general for the two methods, and should therefore 
be self-consistently calibrated. 

There is a significant range in the value of $p_0$ used in the
literature. Reference~\cite{Donato:2008yx} suggests a window between 
66\,MeV and 105\,MeV for the isotropic approach (note that their 
definition of $p_0$ requires a rescaling of $p_0$ by 2 
to compare with our values), while Kadastik 
{\it et al.\/}~\cite{Kadastik:2009ts} 
use a value of $p_0=160$\,MeV for both the isotropic and Monte Carlo 
approach.

\begin{figure}
  \includegraphics[width=0.45\textwidth]{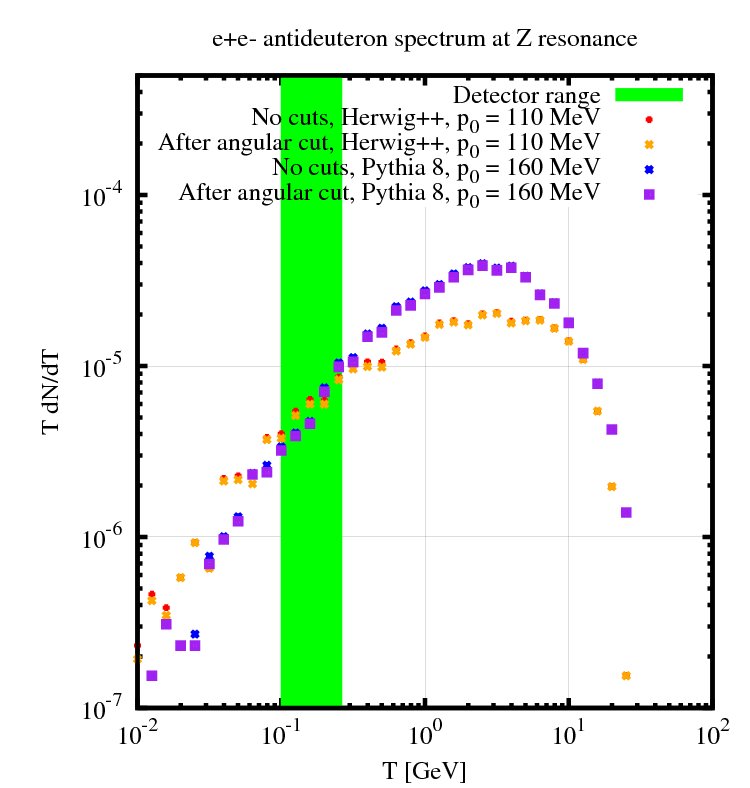}
\caption{Comparison of the energy spectrum $T dN/dT$ of antideuterons for
the calibration reaction, with and without the angular cut $|\cos\theta|<0.95$. The energy range used for the calibration
is shown as the green area. }
\label{calib}
\end{figure}

The relatively large difference of 45\% between the $p_0$ values
found by us using HERWIG++ and  Kadastik {\it et al.\/} using PYTHIA
indicates that the differing hadronization models do indeed lead to
physically different predictions. Since the momentum range used for
the calibration is very small,  our normalization procedure does not
even guarantee that the total number of antideuterons predicted by 
HERWIG++ and PYTHIA for our reference process $e^+e^-\to Z^0\to$~hadrons
agrees. In order to illustrate this point, we repeated the calibration
using PYTHIA (version 8.160). 
In Fig.~\ref{calib}, we  compare the energy spectrum
$TdN/dT$ of antideuterons calculated by us with the two QCD simulations.
By construction, the two spectra cross in the calibration range,
but PYTHIA predicts a harder energy spectrum and a total yield
of antideuterons larger by a factor two than HERWIG.
Going to higher CoM energies or DM masses, one expects moreover that
the spectral shape and the total number of antideuterons evolves differently 
in the two hadronization schemes.
 
Note that the single nucleon spectra calculated with HERWIG++ and
PYTHIA agree quite well, implying that the large differences found in the
antideuteron yield are caused by differences in the two-particle 
correlations of the $\bar p\bar n$ pairs.

%%%%%%%%%%%%%%%%%%%%%%%%%%%%%%%%%%%%%%%%%%%%%%%%%%%%%%%%%%%%%%%%%%%%%%%%%%%%
\subsection{The hadronization models}

Let us recall briefly the main features of the hadronization models employed
in HERWIG++ and PYTHIA: The latter implements the string 
fragmentation or Lund model which is based on a linear confinement 
picture~\cite{lund}.
As supported by lattice QCD studies, this model assumes that the energy stored 
in the colour dipole field between a color charge and an anti-charge increases 
linearly with their separation. As the $q\bar q$ pair moves 
apart from their common production vertex, a color flux tube is stretched 
between them which is described by a massless relativistic string with no 
transverse degrees of freedom. The potential energy stored in the string 
increases as the quarks separate until the string breaks producing a new  
colour-singlet $q^\prime \bar q^\prime$ pair. The string break-up is modeled
as a tunneling process, leading to a flavour-independent Gaussian spectrum 
for the transverse momentum $p_T$.
Baryon production in the Lund model is described similarly as a tunneling
process.

While the Lund model used in  PYTHIA is based on a linear confinement 
picture, the cluster fragmentation model of HERWIG++  assumes ``pre-confinement''
and local parton-hadron duality~\cite{pre}.  At  the end of the perturbative 
QCD cascade,
gluons are splitted non-perturbatively into light quark-antiquark or 
diquark-antidiquark pairs. Color connected quarks are then combined to form 
color singlet 
clusters, using prescribed  mass distributions which fall rapidly for 
large masses. If a cluster is too light to decay, it is replaced by the 
lightest hadron of its flavour, and its mass is shifted to the correct 
value by exchanging momentum with a neighbouring cluster. Heavier clusters fragment
according to empirical prescriptions. For an extensive discussion of
the physics underlying these QCD Monte Carlo simulations we recommend
interested readers to consult Refs.~\cite{lund,ellis}.

\subsection{Uncertainty from the model parameters}

While our main focus is the difference in the antideuteron spectrum between the two hadronization models, 
we also investigate the inherent uncertainty of the HERWIG++ cluster hadronization. 
This uncertainty emerges because ranges of values for the hadronization parameters are compatible with experimental data.
We choose the hadronization parameters
%\footnote{We also include the parameter {\tt AlphaMZ}, which is not a parameter of the hadronization model, but highly relevant nonetheless.} 
and ranges studied in Ref.~\cite{Richardson:2012bn}, as shown in Table~\ref{parameter}.
We run simulations using the ``high'' or ``low'' value for a single parameter,
while keeping the other parameters at their default value.
All simulations were done for the previously discussed case of antideuteron production 
at the $Z$ resonance using $10^8$ events and $p_0=110$~MeV. 
We then calculate the ratios
$$
 R_P\equiv \frac{dN/dT|_{\rm modified}}{dN/dT|_{\rm default}}
$$  
of the antideuteron spectra obtained using modified parameters with respect to the spectra obtained
using default values.
In order to distinguish between changes due to modified two-particle correlations and trivial changes due to modified antinucleon multiplicities, 
we additionally consider the corresponding ratio for antideuteron spectra calculated using Eq.~\eqref{Coal} in the isotropic coalescence approach.
Parameters that affect the two-particle correlations are of particular interest, 
as they could give significant changes in the antideuteron spectrum while keeping the antiproton spectrum in agreement with observed values.
We note that the value of $p_0$ should be adjusted according to the antideuteron yield.
For cases where the ratio is largely independent of energy, this re-calibration can significantly reduce or entirely absorb the effect of adjusting the parameter. 

\begin{figure}
  \includegraphics[width=0.45\textwidth]{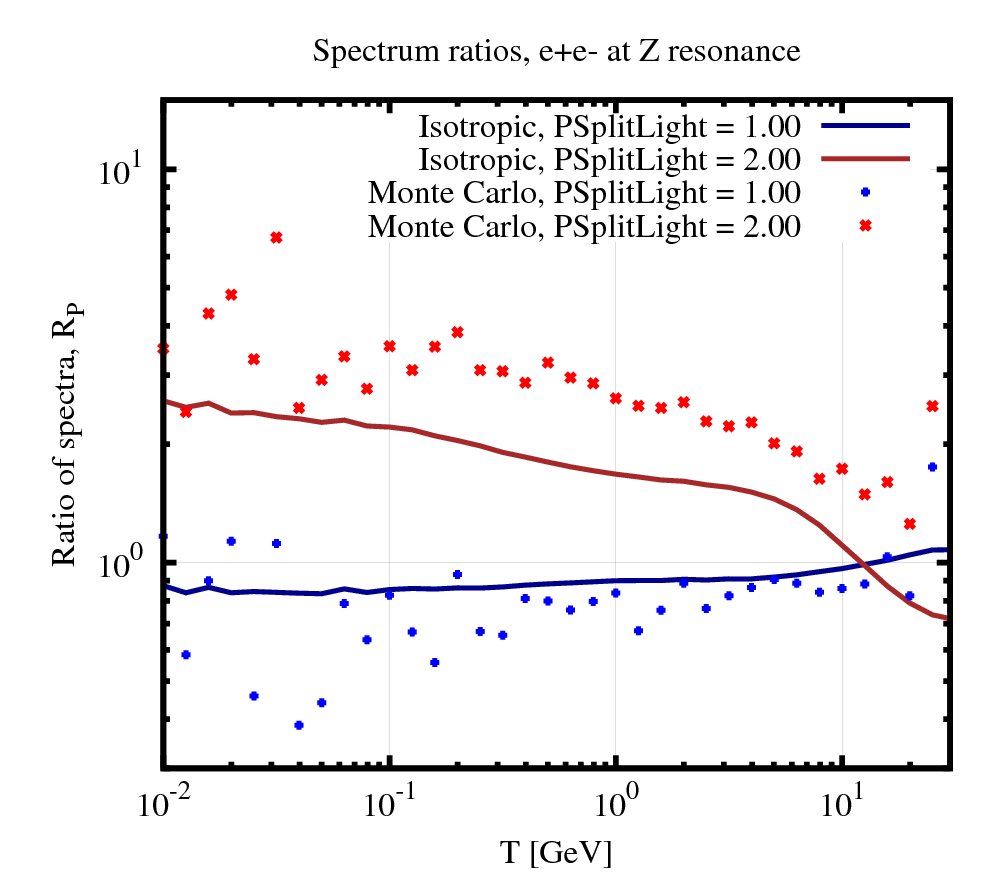}
\caption{Ratio $R_P$ for antideuteron spectra at the $Z$ resonance,
varying the parameter {\tt PSplitLight}.
\label{fig:new1}}
\end{figure}

\begin{figure}
  \includegraphics[width=0.45\textwidth]{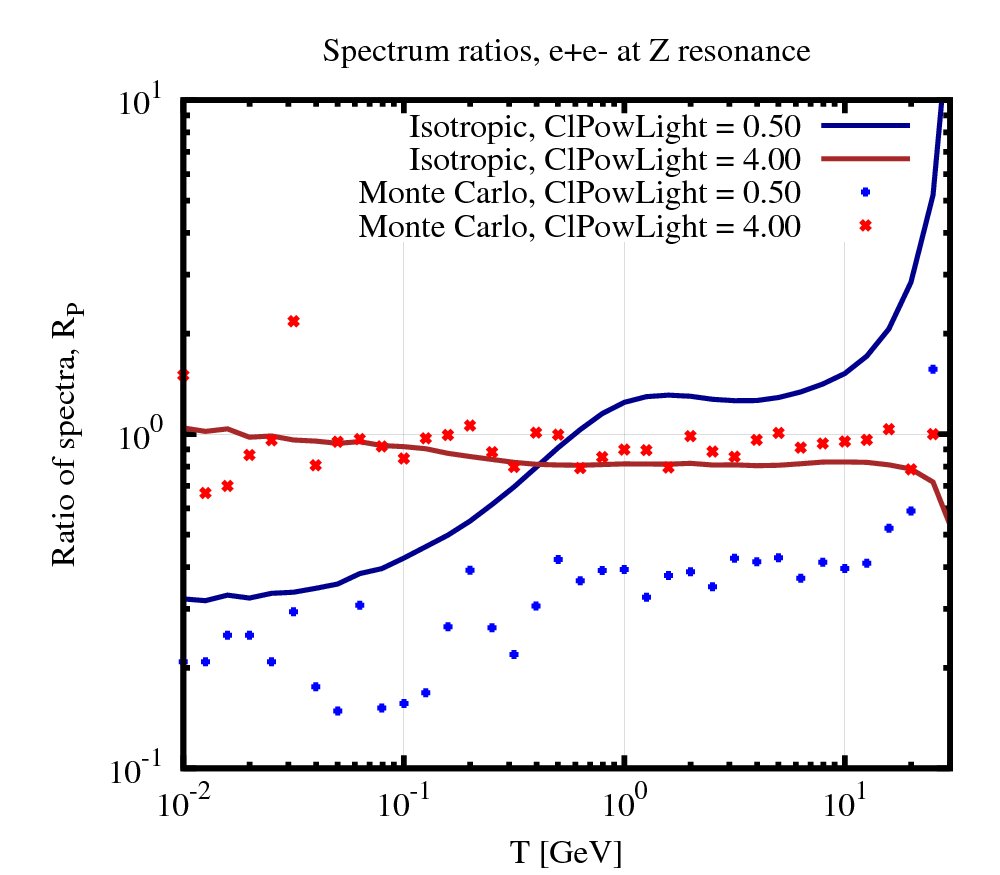}
\caption{Ratio $R_P$ for antideuteron spectra at the $Z$ resonance,
varying the parameter {\tt ClPowLight}.
\label{fig:new2}}
\end{figure}

\begin{figure}
  \includegraphics[width=0.45\textwidth]{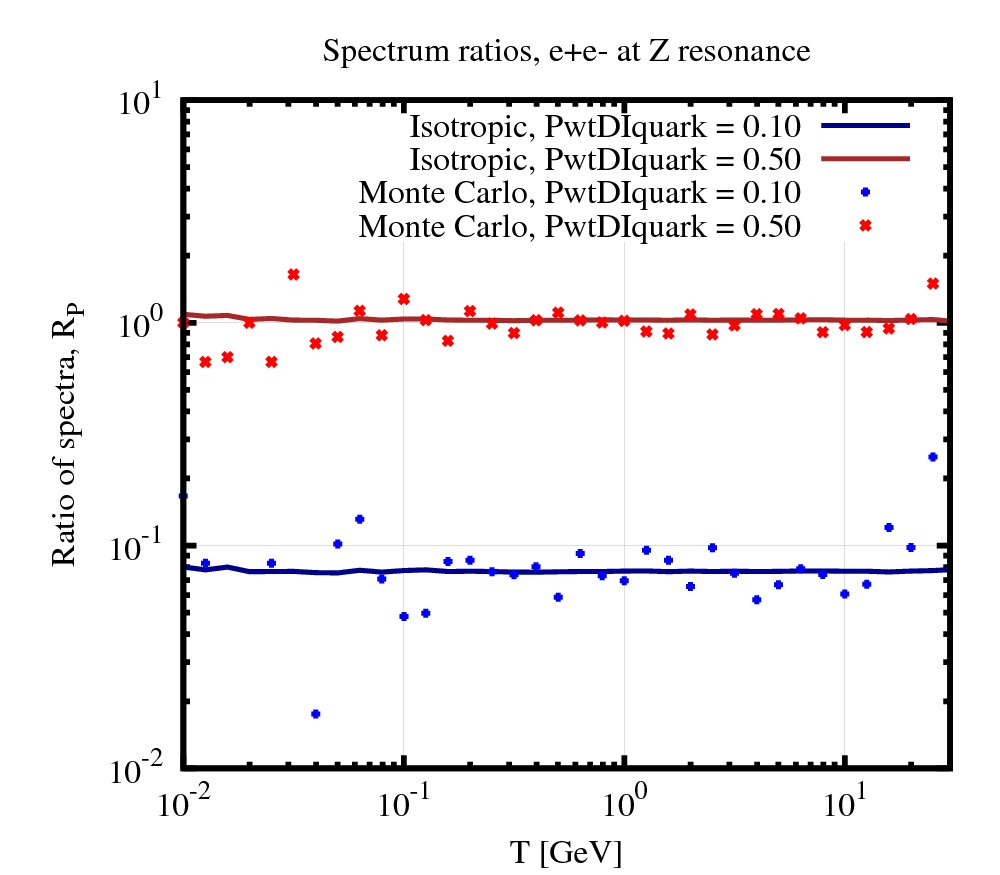}
\caption{Ratio $R_P$ for antideuteron spectra at the $Z$ resonance,
varying the parameter {\tt PwtDIquark}.
\label{fig:new3}}
\end{figure}

In Fig.~\ref{fig:new1}, we show the ratios obtained varying
the parameter {\tt PSplitLight} as an example of a case where the 
antideuteron yield changes non-trivially. While the ratio as a function of energy has a shape similar to that which can be expected from the changes in the antinucleon spectra, 
the change in the overall yield is significantly larger. 
The parameter {\tt ClPowLight} is another case where the antideuteron yield changes non-trivially. 
As seen in Fig.~\ref{fig:new2}, this is, however, a case where also the shape of the spectrum deviates significantly from what can be expected from the changes in the antinucleon spectra.
We find a similar behavior for the parameter {\tt AlphaMZ}, but in this case the non-trivial component is smaller and the ratios are flatter.
As an example of a parameter that influences the antideuteron yield
only via the antinucleon multiplicity, we show the ratios obtained from varying {\tt PwtDIquark} in Fig.~\ref{fig:new3}.
For this and the remaining parameters, the ratio $R_P$ is a rather flat function of energy, and is well represented by the ratio $R_{\rm P,tot}$ of the total antideuteron multiplicities, 
as given in Table~\ref{parameter}.

Overall, we found no case where a change in a parameter of the
cluster fragmentation model strongly influenced the antideuteron spectra
without also leading to a significant change in the nucleon spectra.
Moreover, a re-calibration of $p_0$ will in most cases contribute to reduce changes in the antideuteron spectrum from adjustments of hadronization parameters.
We therefore conclude that the uncertainties within the hadronization model used by HERWIG++ does not lead to a spread in the
predicted antideuteron yield that is significantly larger than the factor of a few difference between HERWIG++ and PYTHIA.

\begin{table}
\begin{tabular}{l|c|c|c|c|c}
Parameter & ``low'' & $R_{\rm P,tot}^{\rm low}$ & default & ``high'' & $R_{\rm P,tot}^{\rm high}$ \\ \hline
{\tt GluonMass}   & 0.75 & 0.85 & 0.95 & 1.00 & 0.81 \\
{\tt AlphaMZ}     & 0.10 & 1.62 & 0.12 & 0.12 & 1.00 \\
{\tt ClPowLight}  & 0.50 & 0.36 & 1.28 & 4.00 & 0.93 \\
{\tt ClMaxLight}  & 3.00 & 0.97 & 3.25 & 4.20 & 0.84 \\
{\tt ClSmrLight}  & 0.30 & 1.04 & 0.78 & 3.00 & 0.96 \\
{\tt PSplitLight} & 1.00 & 0.81 & 1.14 & 2.00 & 2.49 \\
{\tt PwtDIquark}  & 0.10 & 0.08 & 0.49 & 0.50 & 1.00 \\
{\tt PwtSquark}   & 0.50 & 0.99 & 0.68 & 0.80 & 1.01
\end{tabular}
\caption{\label{parameter}
Varied parameters of HERWIG++ with the corresponding ``low'', default and ``high'' 
values and the resulting ratios $R_{\rm P,tot}$ of total antideuteron multiplicities obtained in the
event-by-event Monte Carlo approach.}
\end{table}

%%%%%%%%%%%%%%%%%%%%%%%%%%%%%%%%%%%%%%%%%%%%%%%%%%%%%%%%%%%%%%%%%%%%%%%%%%%%%%%%%%%%%%%%%%%%%%%%%%%%%%%%%%%%%%%%%%
%%%%%%%%%%%%%%%%%%%%%%%%%%%%   SECTION: Computational results: Source spectra   %%%%%%%%%%%%%%%%%%%%%%%%%%%%%%%%%%
%%%%%%%%%%%%%%%%%%%%%%%%%%%%%%%%%%%%%%%%%%%%%%%%%%%%%%%%%%%%%%%%%%%%%%%%%%%%%%%%%%%%%%%%%%%%%%%%%%%%%%%%%%%%%%%%%% 

\section{DM antideuteron spectra} \label{sec:source}

Our results were generated using $10^8$ events for each annihilation channel
and dark matter mass, and are generally presented in terms of the scaled 
kinetic energy 
$x\equiv T/M_{\rm DM}$, where $M_{\rm DM}$ is the dark matter mass. For practical
reasons, we used the lightest neutralino in the MSSM as DM particle,
thus allowing us to use MadGraph~\cite{mad} to generate events to be fed into HERWIG++ for 
showering and hadronization.
Note that while the branching ratios into different final states are strongly
model-dependent, the energy spectrum of a specific final-state is---except 
in cases where spin correlations are important---mainly determined
by the CoM energy $\sqrt{s}=2M_{\rm DM}$.

\subsection{The antideuteron fragmentation spectra}

\begin{figure}
  \includegraphics[width=0.45\textwidth]{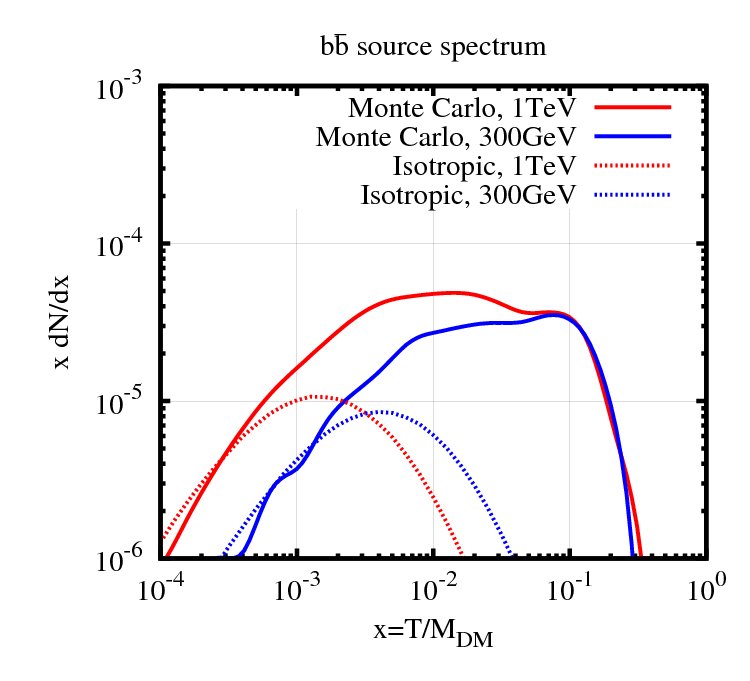}
  \hfil
  \includegraphics[width=0.45\textwidth]{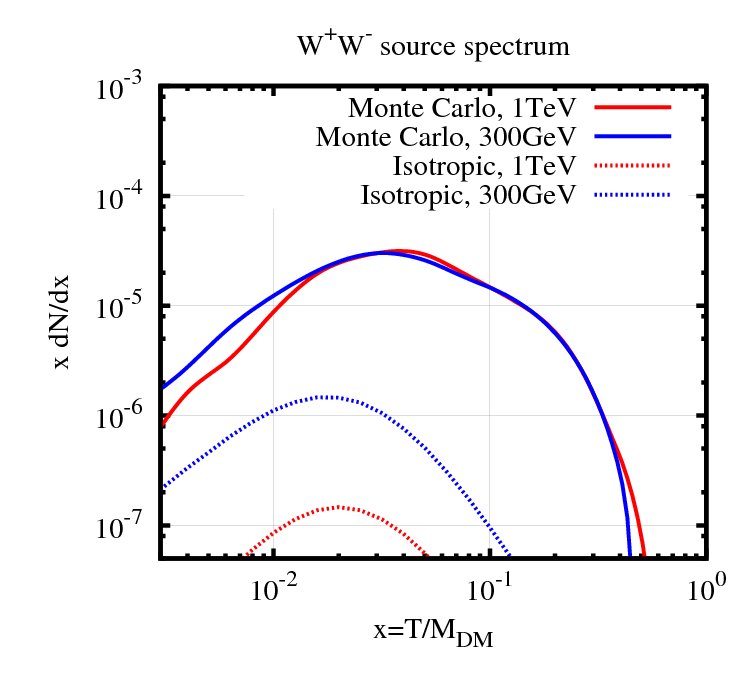}  
\caption{Antideuteron source spectra for dark matter annihilations 
into $b\bar{b}$ (top) and $W^+W^-$ (bottom). 
The solid lines show the spectra for per-event coalescence within the Monte Carlo, while the dotted lines show the spectra for coalescence of the average antiproton and antineutron spectra. Red lines show the result for a dark matter mass of 1 TeV, blue lines for 300\,GeV.}
\label{fig:bbWWspectra}
\end{figure}

The antideuteron fluxes from the isotropic and Monte Carlo approaches 
obtained by us using HERWIG++ are 
plotted for DM masses of 300\,GeV and 1\,TeV in Fig.~\ref{fig:bbWWspectra}. 
In the $b\bar{b}$ case, the low-$x$ part of the energy spectra $dN/dx$ are roughly 
the same for the two methods. The spectra in the Monte Carlo approach 
extend, however, towards much larger $x$. For the $W^+W^-$ case, there is a 
significant difference in the antideuteron spectrum for all $x$
between the two approaches. For 1\,TeV, the difference is of order $100$ at low $x$, 
while it is around a factor $\sim 30$ for 300\,GeV. 
The shapes of the spectra also differ somewhat between the two approaches, 
with the maxima of $xdN/dx$  shifted towards higher $x$ in the Monte Carlo
approach.

It is clear that when using the correct Monte Carlo approach, the difference in magnitude between the $b\bar{b}$ and $W^+W^-$ antideuteron spectra becomes much smaller than in the isotropic approach. 
Another interesting feature is that this difference appears to depend on the DM mass.

\begin{figure}[h]
	\centering
	\includegraphics[width=0.45\textwidth]{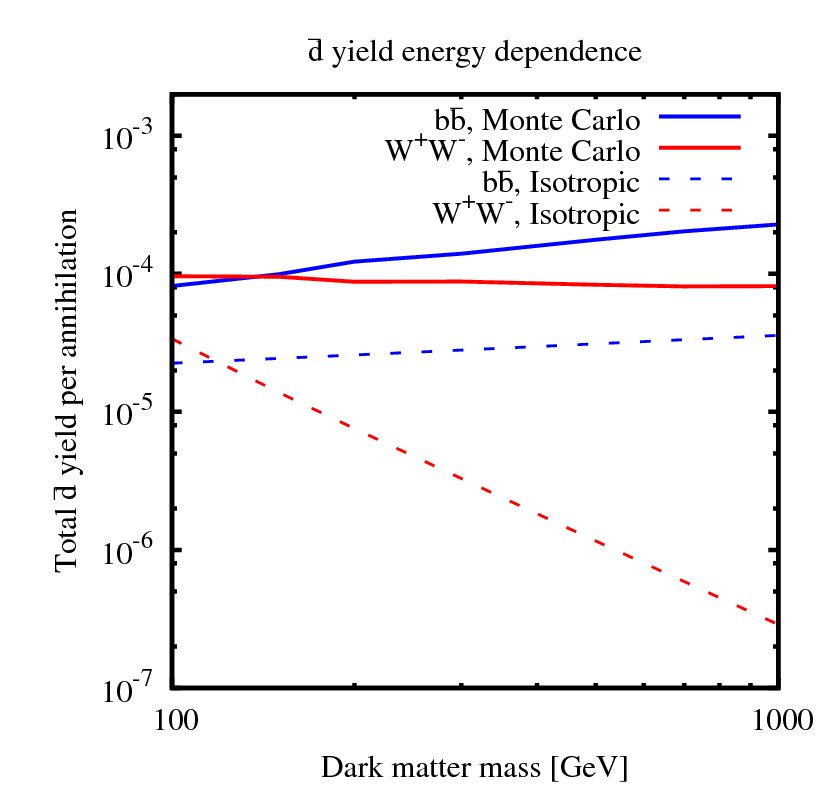}
	\caption{Average total antideuteron yield per DM annihilation event. The solid lines show the results using per-event coalescence, while the dashed lines show the results using the isotropic approximation.}
	\label{fig:num_evo}
\end{figure}

In order to investigate the dependence of the antideuteron spectra on 
the DM mass, we plot the total number of antideuterons produced against 
the DM mass for the two approaches in Fig.~\ref{fig:num_evo}.  Comparing 
this graph to the corresponding one of Ref.~\cite{Kadastik:2009ts} obtained 
using PYTHIA, we find that the general behavior of the antideuteron yields as
function of channel and approach agrees well. Comparing individual energy
spectra, we observe a sharper drop at high energies in the spectra produced 
by HERWIG++. We also note some differences in the overall magnitudes, and we 
therefore proceed by comparing our HERWIG++ results to those of Kadastik 
{\it et al.\/}~\cite{Kadastik:2009ts} in more detail. 
For their results we use the data files (without EW corrections)
from Ref.~\cite{Cirelli:2010xx}, which were generated using PYTHIA version 8.135.

\begin{figure}[h]
	\centering
	\includegraphics[width=0.45\textwidth]{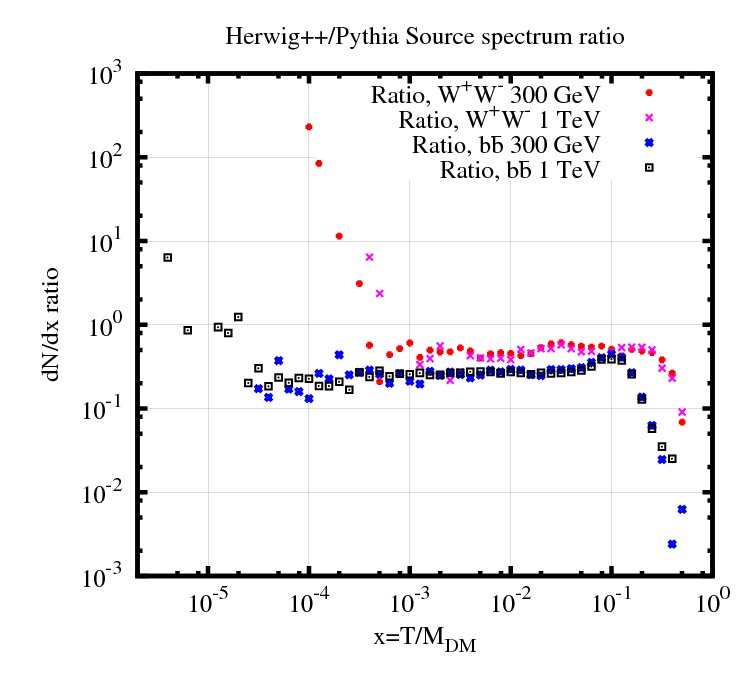}
	\caption{Ratio $R$ of the antideuteron spectra predicted by
HERWIG++ and PYTHIA as function of $x$ for neutralino annihilations
with mass 300\,GeV and 1\,TeV, respectively.}
	\label{fig:ratio}
\end{figure}

\begin{figure}
  \includegraphics[width=0.45\textwidth]{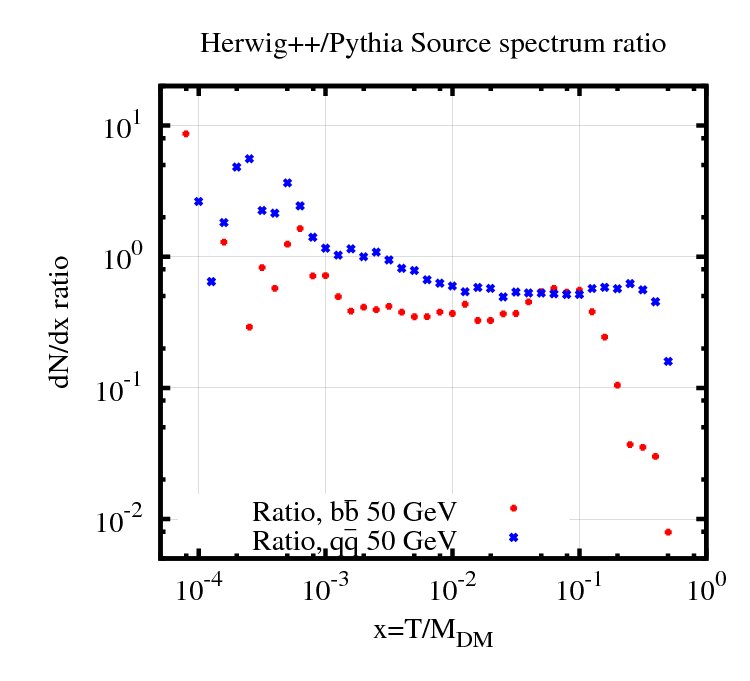}
\caption{Ratio $R$ of the antideuteron spectra predicted by
HERWIG++ and PYTHIA as function of $x$ for
$\chi\chi\to b\bar b$ and $\chi\chi\to q\bar q$ with $m_\chi=50$\,GeV and $q=u,d,s$.}
\label{fig:nnqq}
\end{figure}

In Fig.~\ref{fig:ratio}, we show the ratio
$$
 R\equiv \frac{dN/dx|_{\rm HERWIG++}}{dN/dx|_{\rm PYTHIA}}
$$ 
of the antideuteron energy spectra calculated with HERWIG++ and PYTHIA,
respectively. In the peak region, this ratio is close to 0.5 for the 
$W^+W^-$ channel, while it is close to 0.3 for the 
$b\bar b$ channel. In the forward region $x\to 1$, PYTHIA predicts
a significantly higher antideuteron yield than HERWIG++, leading a 
drop of the ratio $R$. 
As expected from the calibration reaction, cf.~Fig.~\ref{calib}, 
 the ratio $R$ lies below one in the peak region,
followed by a crossover to $R>1$ at lower $x$.
The expected crossover is not seen in the 300\,GeV quark case, possibly due to insufficient statistics at low $x$.
For the gauge bosons, the $x$ value for which the ratio crosses one appears to be shifted slightly toward higher values with increasing DM mass, but the uncertainty with $10^8$ events is too large to determine the crossover points precisely.
For the quark case, the crossing region is shifted towards lower $x$. 
In order to see if there is a DM mass dependence of the crossing point in the quark channel, we also investigate the ratio at $M_{\rm DM}=50$\,GeV, as seen in Fig.~\ref{fig:nnqq}.
In addition to the $b$-quarks studied in the rest of the article, the figure also includes the corresponding ratio for annihilations into light quarks ($u,d,s$).
We see that not only does the crossover for the $b$-quark case appear at higher $x$ for lower $M_{\rm DM}$, but the 
 ratios of the antideuteron energy spectra in general also differ somewhat between light and heavy quarks.

%%%%%%%%%%%%%%%%%%%%%%%%%%%%%%%%%%%%%%%%%%%%%%%%%%%%%%%%%%%%%%%%%%%%%%%%%%%%%
\subsection{Interpretation} 

The perhaps most striking feature of Fig.~\ref{fig:num_evo} is how strongly 
the impact of going from the isotropic to the Monte Carlo approach varies
between the $b\bar b$ and $W^+W^-$ channel. In the latter case, changing
the DM mass by a factor 10 leads to a factor 100 difference in the
antideuteron yield predicted by the two approaches, while in the
former case the prediction differs only by a factor three.
This suggests that the basic assumption underlying the ``isotropic approach,''
namely an uncorrelated isotropic emission of the final-state particles,
is not too strongly violated for $q\bar q$ final-states.

\begin{figure*}
  \includegraphics[width=0.45\textwidth]{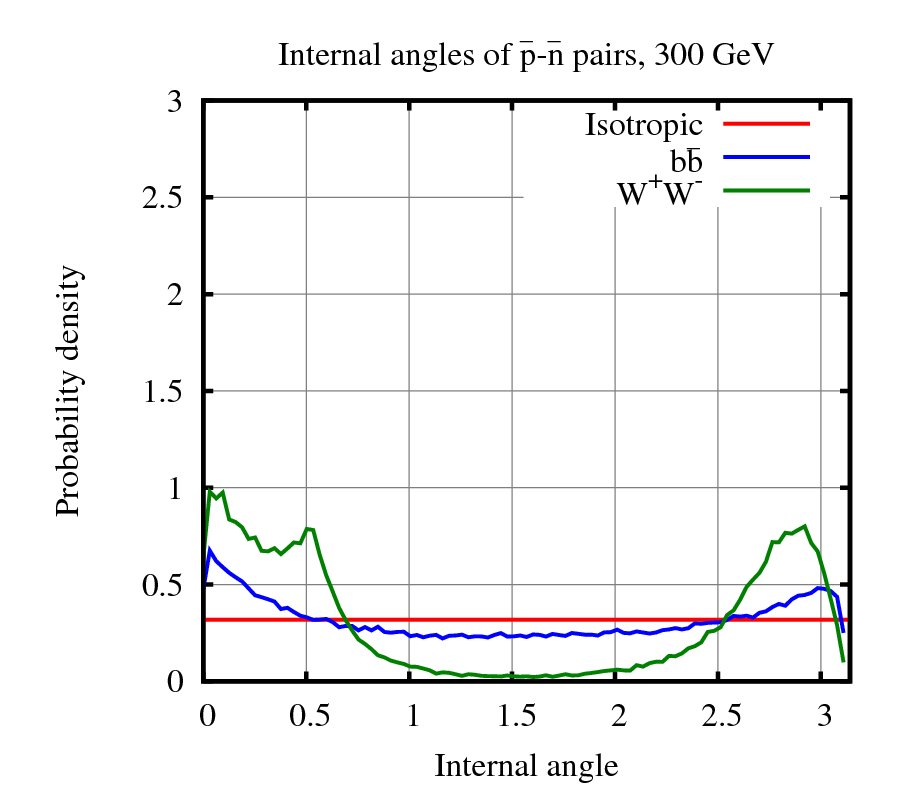}
  \hfil
  \includegraphics[width=0.45\textwidth]{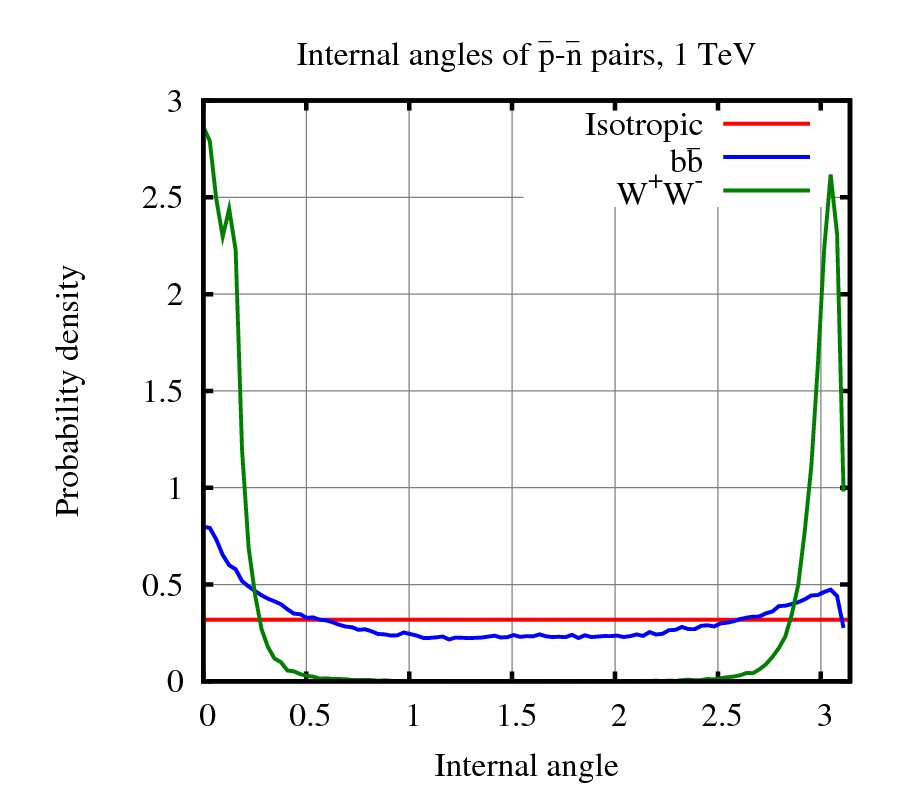}           
\caption{Distribution of the angles between the momenta of antiprotons and antineutrons; left for $M_{\rm DM}=300$\,GeV, right for $M_{\rm DM}=1$\,TeV. Blue and green lines show the distribution in ``Monte Carlo'' approach result for $b\bar{b}$ and $W^+W^-$, respectively, while the red lines show the result of an isotropic distribution for comparison.}
\label{fig:Angl}
\end{figure*}

To illustrate this effect, we show in Fig.~\ref{fig:Angl} the distribution 
of the angles between the momenta of antiproton and antineutron pairs. 
These distributions are strongly peaked for angles near 0 and $\pi$  
for the $W^+W^-$ channel, while the peaks are much less pronounced
for  $b\bar{b}$. More importantly, in the quark channel the shape of the 
distribution changes only mildly when increasing the DM mass, while the peaks
in the $W^+W^-$ channel increase strongly.

\begin{figure}[h]
	\centering
	\includegraphics[width=0.45\textwidth]{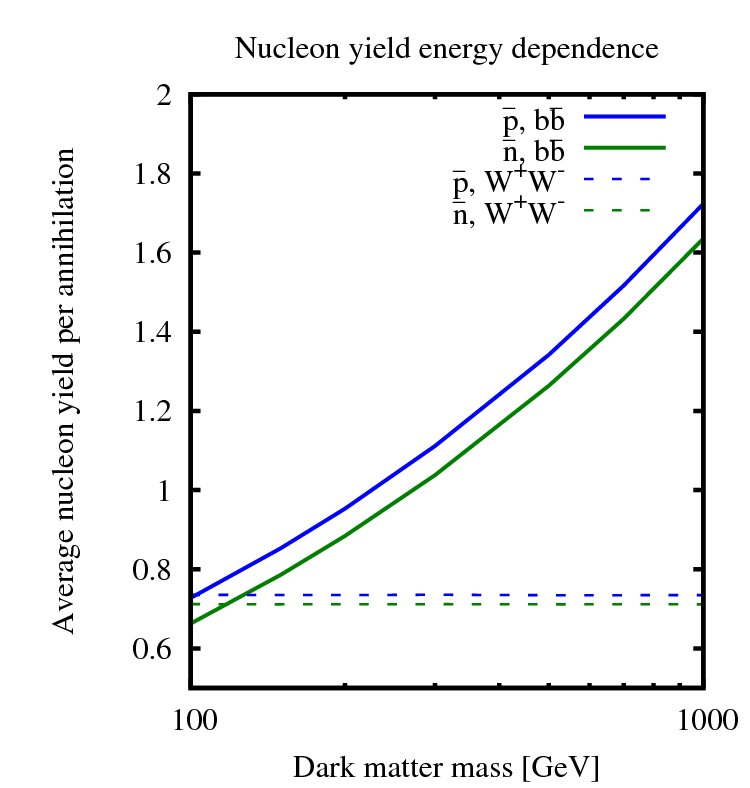}
	\caption{Average total antiproton and antineutron yields per annihilation event as function of the dark matter mass. Green indicates antineutrons, while blue indicates antiprotons. The solid lines show the results for the $b\bar{b}$ case, while the dashed lines show the results for the $W^+W^-$ case.}
	\label{fig:nucleon_evo}
\end{figure}

The latter effect is a simple reflection of relativistic beaming:
The $W$'s fed into HERWIG++ from MadGraph are decaying on-shell, and their decay product are 
emitted in a cone of opening angle $\theta\sim m_W/M_{\rm DM}$.
The relativistic beaming in itself does not affect coalescence in the Monte Carlo approach, as the nucleon momenta are to be evaluated in the CoM frame of the respective $\bar p\bar n$ pairs. 
The fact that the bosons are decaying on-shell, however, implies that the nucleon multiplicity---and thus the antideuteron multiplicity---remains constant with increasing $M_{\rm DM}$, 
as shown in Fig.~\ref{fig:nucleon_evo}.
Using the (wrong) Eq.~\eqref{Coal} derived in the isotropic approach,
\be
  \frac{dN_{\bar{d}}}{dx} \propto \frac{1}{M_{\rm DM}^2} 
  \frac{dN_{\bar{n}}}{dx} \frac{dN_{\bar{p}}}{dx} ,
\ee
one would instead expect an $1/M_{\rm DM}^2$ suppression in the antideuteron yield, as is reflected by the isotropic $W^+W^-$ result in Fig.~\ref{fig:num_evo}.

In contrast to the gauge bosons,
the $b$'s are treated as virtual particles and start QCD cascades.
Although these cascades are angular-ordered and lead to confined jets,
the overall shape of the event is not too far from spherically symmetric.
Moreover, the two initial back-to-back jets are color connected, and partons from the two jets have to combine 
in order to produce colorless final states.

The fact that the antideuteron yield does not show
the  $1/M_{\rm DM}^2$ suppression in the $b\bar{b}$ channel using the
isotropic approach can also be explained. In this case, the growth of the nucleon multiplicity
shown in Fig.~\ref{fig:nucleon_evo} with $M_{\rm DM}$ overcompensates 
the $1/M_{\rm DM}^2$ factor, leading to a slight increase of the antideuteron
yield even in the isotropic approach.

Finally, we discuss the ratios of the HERWIG++ and PYTHIA fragmentation spectra.
As already mentioned, the $x$ for which $R$ crosses one in the $W^+W^-$ channel appears to shift slightly towards higher $x$ with increasing DM mass: 
As the gauge bosons decay on-shell, their decay products are boosted according to the DM mass, and for $M_{\rm DM} \gg m_{W}$ we expect $T_{\bar d}$ to increase linearly with $M_{\rm DM}$.
With the $T$-value of the crossover point increasing linearly with $M_{\rm DM}$, 
the corresponding $x=T/M_{\rm DM}$ should then converge towards a constant value with increasing DM masses.

In the $b\bar b$ channel, however, we see from Figs.~\ref{fig:ratio} and~\ref{fig:nnqq} that the $x$-value of the crossing region appears to decrease as $\sim1/M_{\rm DM}$, 
such that the crossing occurs at a roughly constant value of the antideuteron kinetic energy $T_{\bar d}$. 
Note, however, that the drop in the ratio for $x\to 1$ appears at a constant value of $x$ rather than $T$.

Using the differences between HERWIG++ and PYTHIA as an estimate
for the uncertainty introduced by the hadronization models, we conclude
that the high-energy part of the antideuteron spectra cannot
be predicted reliably for neither the gauge boson nor quark channels.
For the $b \bar b$ channel, the spectrum has an uncertainty of a factor $\sim 3$ from $x\sim 10^{-1}$ down to $T\sim 10^{-2}$\,GeV, below which the uncertainty increases.
For the gauge bosons, we find an uncertainty of $\sim2$ for $10^{-3}\lesssim x \lesssim 0.5$, with rapidly increasing uncertainties outside this range.
For the $W^+W^-$ channel, the crossover point, and consequently the high uncertainty region at low energies, moves linearly towards higher $T$ with increasing DM mass.
Since $T$, rather than $x$ is the relevant quantity for the observable intensity of antideuterons, this implies that observationally 
relevant energies could lie in the  region with large uncertainties for DM masses in the TeV range and above.

%%%%%%%%%%%%%%%%%%%%%%%%%%%%%%%%%%%%%%%%%%%%%%%%%%%%%%%%%%%%%%%%%%%%%%%%%%%%
\subsection{Higher order processes}

Motivated by the PAMELA excess, previous works have discussed the 
annihilation and the resulting antideuteron fluxes of DM particles 
up to 30\,TeV~\cite{Kadastik:2009ts,Brauninger:2009pe}. The unitarity limit of 
50\,TeV
implies that the partial wave amplitude for the annihilation of such a
heavy DM particles is close to one. Thus DM particles with masses 
$\gsim 10$\,TeV should behave as {\em strongly\/} interacting particles,
and higher order processes should give an important contribution
to their total annihilation cross section.

\begin{figure}[h]
	\centering
	\includegraphics[width=0.5\textwidth]{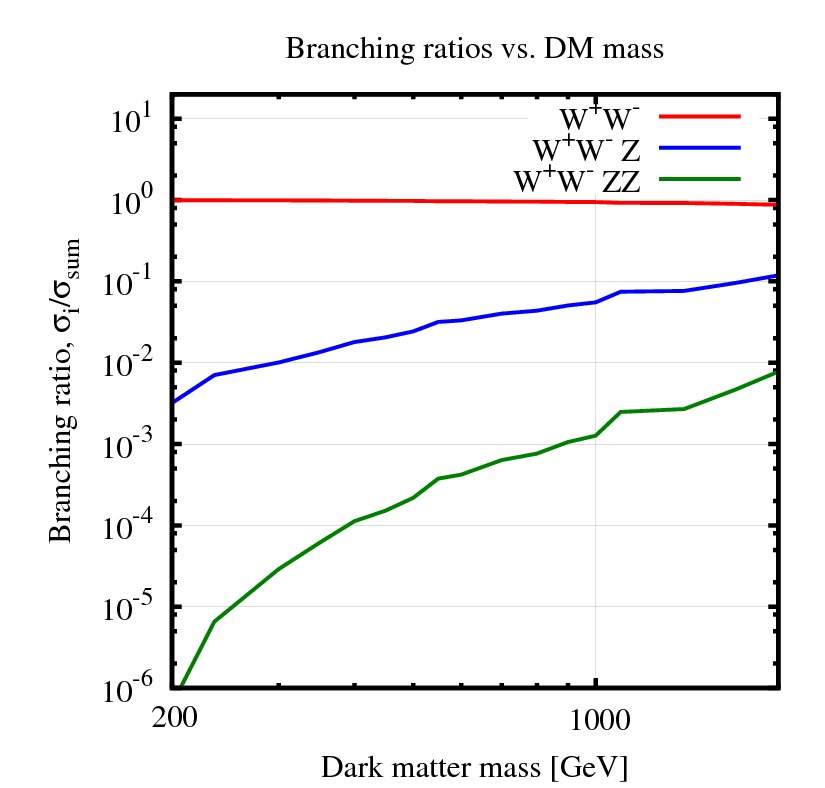}
	\caption{Branching ratios for various annihilation channels as function of the dark matter (neutralino) mass. The red line shows the benchmark branching ratio for the tree level process $\tilde{\chi}^0_1 \tilde{\chi}^0_1 \rightarrow W^+W^-$. The blue and green lines show the branching ratios for the higher order processes $\tilde{\chi}^0_1 \tilde{\chi}^0_1 \rightarrow W^+W^-Z$ and $\tilde{\chi}^0_1 \tilde{\chi}^0_1 \rightarrow W^+W^-ZZ$, respectively.}
	\label{fig:XS_evo}
\end{figure}

As a test of this conjecture,  we calculated the annihilation cross sections 
for $\tilde{\chi}^0_1 \tilde{\chi}^0_1 \rightarrow W^+W^-$, and 
for the higher order tree-level processes in which one and two extra 
$Z$-bosons are emitted.
The calculations were performed for several different neutralino masses in 
the range 200 GeV to 2 TeV.  Using these cross sections, we calculated the 
branching ratios for these processes, normalized to a sum of 1.
The results from these calculations are plotted 
in Fig.~\ref{fig:XS_evo}.

Figure~\ref{fig:XS_evo} shows as expected that the relative contribution 
from higher order processes is negligible for low neutralino masses. The 
contributions do, however, increase rapidly with increasing masses. For 2\,TeV,
 the branching ratio for $W^+W^-Z$ is roughly 10\% of that for the tree level 
process. While the process involving two additional $Z$-bosons is more 
strongly suppressed for low neutralino masses than the single $Z$-boson 
process, 
its relative contribution increases faster with increasing masses than for 
$W^+W^-Z$. 

Extrapolating our results, we expect the contribution from higher order 
processes to become significant for a neutralino mass in the 10\,TeV range
\cite{ewbrems}. 
The emission of additional $Z$-bosons are, of course, not the only possible 
higher order processes, and when performing calculations in the multi TeV 
range, the contributions from a number of processes should be considered.
 
Note also that the results of this subsection are more model-dependent
than the previous ones.
For DM models other than the MSSM, the DM mass  where final-states 
with $n>2$ particles  become important may be lower.

%%%%%%%%%%%%%%%%%%%%%%%%%%%%%%%%%%%%%%%%%%%%%%%%%%%%%%%%%%%%%%%%%%%%%%%%%%%%
\section{Antideuteron intensity at Earth}
 
We now move from the antideuteron spectra produced in a single annihilation 
to the calculation of the resulting antideuteron intensity at Earth. As
charged particles scatter on fluctuations of the Galactic magnetic field
with variation scales comparable to their Larmor radius, their propagation 
resembles a random walk and is well described by the diffusion approximation.
We model this random walk using the so-called two-zone propagation model for the Galaxy.
This is a cylindrical diffusion model consisting of a magnetic halo and a thin gaseous 
disk of radius $R=20$~kpc and half-heights $L$ (a free parameter) and $h=100$~pc, respectively.  
For the numerical values of the parameters in this model, we adopt the three sets presented in \cite{Donato:2003xg} to yield maximal, median and minimal antiproton fluxes from DM annihilations while being compatible 
with the observed B/C ratio.
These sets are labeled `max', `med', and `min', respectively, and listed in 
Table~\ref{tab:PropModels}.
Assuming steady state conditions, the diffusion equation neglecting
reacceleration is given by
\begin{equation} \label{eq:Diffusion}
 -D(T) \nabla^2 f  +  \frac{\partial}{\partial z} ( {\rm sign}(z) f V_c) = 
 Q - 2 h \delta (z) \Gamma_{\rm ann}(T) f \,,
\end{equation}
where $f(\vec{x},T)=dN_{\bar{d}}/dT$ is the number density of antideuterons 
per unit kinetic energy, $D(T)= D_0 \beta \mathcal{R}^\delta$  the 
(spatial) diffusion coefficient, $V_c$ a convective wind perpendicular to the Galactic disk, $z$ the vertical coordinate, $\beta=v/c$ the velocity and
$\mathcal{R}$ the rigidity of antideuterons.

%{\bf break in diffusion?}

\begin{table}
\begin{tabular}{|l|c|c|c|c|}
  \hline
	Model 	& $L$ in kpc 	& $\delta$ 	& $D_0$ in kpc$^2$\,Myr$^{-1}$ 	& $V_c$ in km\, s$^{-1}$	\\ 
    \hline
    	max 	& 15 		& 0.46 		& 0.0765				& 5				\\
    	med	& 4		& 0.7		& 0.0112				& 12				\\
	min	& 1		& 0.85		& 0.0016				& 13.5				\\
    \hline
\end{tabular}
\caption{Propagation parameters for the max, med and min models.}
\label{tab:PropModels}
\end{table}

\begin{figure}
\includegraphics[width=0.5\textwidth]{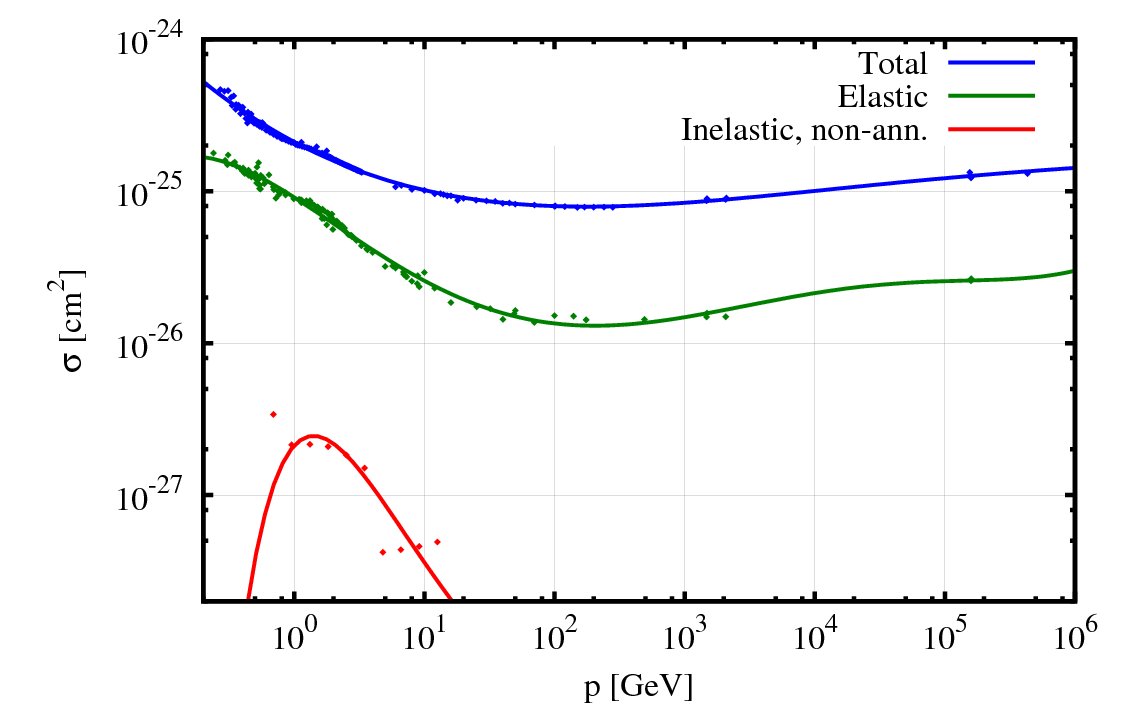}
\caption{Cross section data for antideuterons on interstellar protons as a function of the antideuteron momentum. The points indicate experimental data, while the lines show the fits to the data which were used in our calculations.}
\label{fig:ISM_XS}
\end{figure}

The term $\Gamma_{\rm ann}(T)$ in Eq.~\eqref{eq:Diffusion} accounts for 
annihilations of antideuterons on the interstellar medium.  The annihilation 
rate is given by
\begin{equation}
\Gamma_{\rm ann}  = (n_{H} + 4^{\frac{2}{3}} n_{\rm {He}}) 
                  \langle \sigma^{\rm {ann}}_{\bar{d}p}v\rangle\, ,
\end{equation}
where the factor $4^{\frac{2}{3}}$ accounts for the different cross sections 
of H and He interactions assuming simple geometrical scaling, and we use
$n_{\rm H}\approx 1$\,cm$^{-3}$ and $n_{\rm He} \approx 0.07 n_{\rm H}$ 
as the number density of hydrogen and helium in the disk, respectively.
The fit to the experimental data~\cite{Amsler:2008zzb,Schopper1988} we used for these 
reactions is shown in Fig.~\ref{fig:ISM_XS}.

Finally, the source term $Q$ is fixed by the DM halo profile, which we choose
as either a Navarro-Frenk-White (NFW) profile~\cite{Navarro:1995iw}
($\alpha = 1$, $\beta = 3$, $\gamma =1$) or an 
isothermal profile ($\alpha = 2$, $\beta = 2$, $\gamma =0$) in
\begin{equation}
 \rho(r) = \frac{\rho_0}{(r/a)^{\gamma}\left[ 1 + \left(r/a \right)^{\alpha} \right]^{(\beta - \gamma)/\alpha}} \,,
\end{equation}
or as the Einasto profile,
\begin{equation}
\rho_{\rm Einasto}(r) = \rho_0 \exp \left[-\frac{2}{\alpha} \left( \left( \frac{r}{a} \right)^{\alpha} - 1 \right) \right] , \ \ \  \alpha =0.17 .
\end{equation}
Values of the free parameters $\rho_0$ and $a$ in these profiles are given for 
the Milky Way in Table~\ref{tab:DensityParam}.

\begin{table}
\begin{tabular}{|l|c|c|}
  \hline
	DM Halo Profile &	$\rho_0$ in GeV/cm$^3$ &	$a$ in kpc \\ 
    \hline
	NFW & 			0.26 & 				20 \\
	Isothermal & 		1.16 & 				5 \\
	Einasto &		0.06 &				20 \\
    \hline
\end{tabular}
\caption{Density profile parameters.}
\label{tab:DensityParam}
\end{table}

\begin{figure}[h]
  \includegraphics[width=0.45\textwidth]{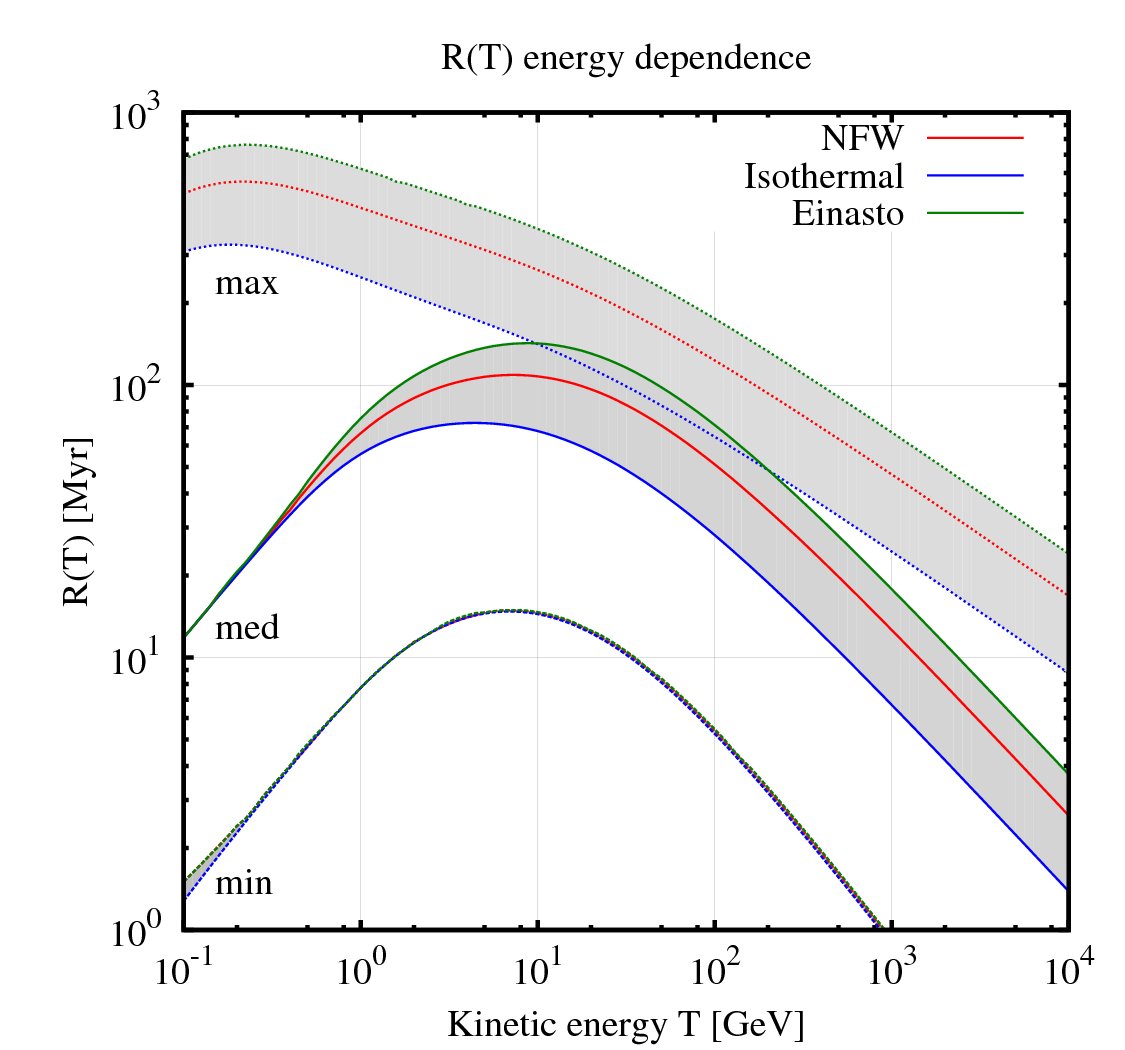}
  \caption{The function $R(T)$, plotted for different dark matter profiles and propagation settings. The filled grey areas show the differences in $R(T)$ between the density profiles for a given propagation model. The upper lines correspond to the `max' model, the middle lines correspond to the `med' model, and the lower lines correspond to the `min' model.\label{fig:R_T}}
\end{figure}

The solution to the diffusion equation Eq.~\eqref{eq:Diffusion} gives the 
intensity $I_{\bar{d}}$ at the  position of the Earth as 
\cite{Donato:2003xg,Maurin:2001sj}
\begin{equation} \label{eq:FluxEarth}
 I_{\bar{d}}(T,\vec{r}_{\odot}) = 
 B \frac{v_{\bar{d}}}{4 \pi} \left( \frac{\rho_{0}}{M_{\rm DM}} \right)^2 R(T)
 \frac{ \left<\sigma v \right> }{2} \frac{dN_{\bar{d}}}{dT}  \,,
\end{equation}
where $B$ is a possible boost factor,
and the astrophysics is encoded in the propagation function $R(T)$.
Our results for $R(T)$ are shown in Fig.~\ref{fig:R_T} and  agree  well 
with those of Kadastik {\it et. al.} \cite{Kadastik:2009ts} (when exchanging the labels 
for the Einasto and Moore profiles in their figure).

Combining the propagation function $R(T)$ with the calculated energy
spectra of antideuterons and accounting for solar modulations by
replacing $T$ with $T_\odot=T-|Ze|\phi_{\rm Fisk}$ with 
$\phi_{\rm Fisk}=0.5$\,GV~\cite{Gleeson:1968zza}, we are in position to determine the
intensity of antideuterons at the position of the
Earth. For the final results, we adopt for the thermally averaged cross section 
$\langle\sigma v \rangle = 3 \times 10^{-26}{\rm cm^3/s}$, a boost 
factor $B=1$ for each annihilation channel, and use the NFW density profile. 

\begin{figure*}
  \includegraphics[width=0.45\textwidth]{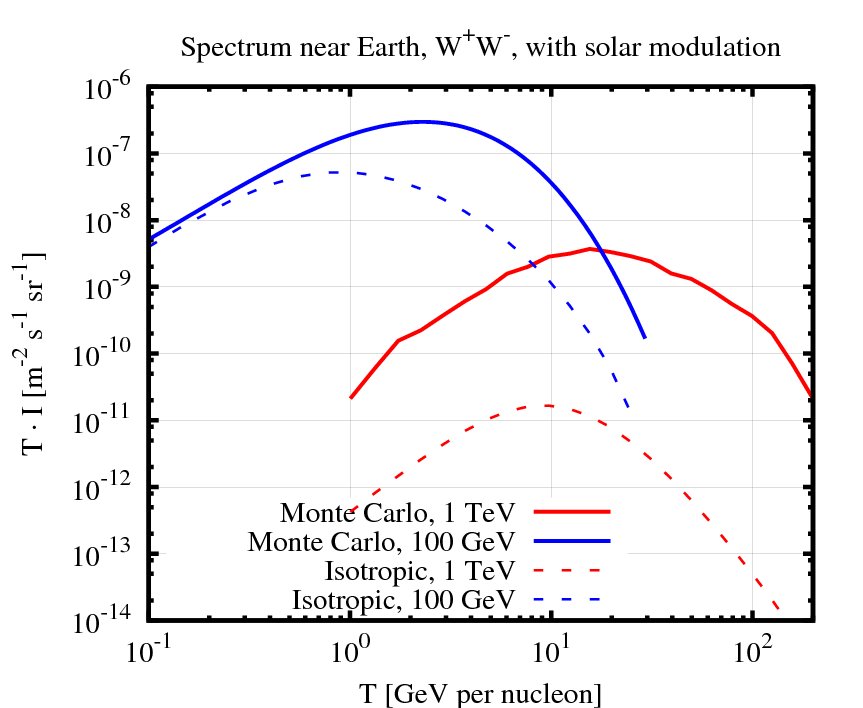}              
  \hfil
  \includegraphics[width=0.45\textwidth]{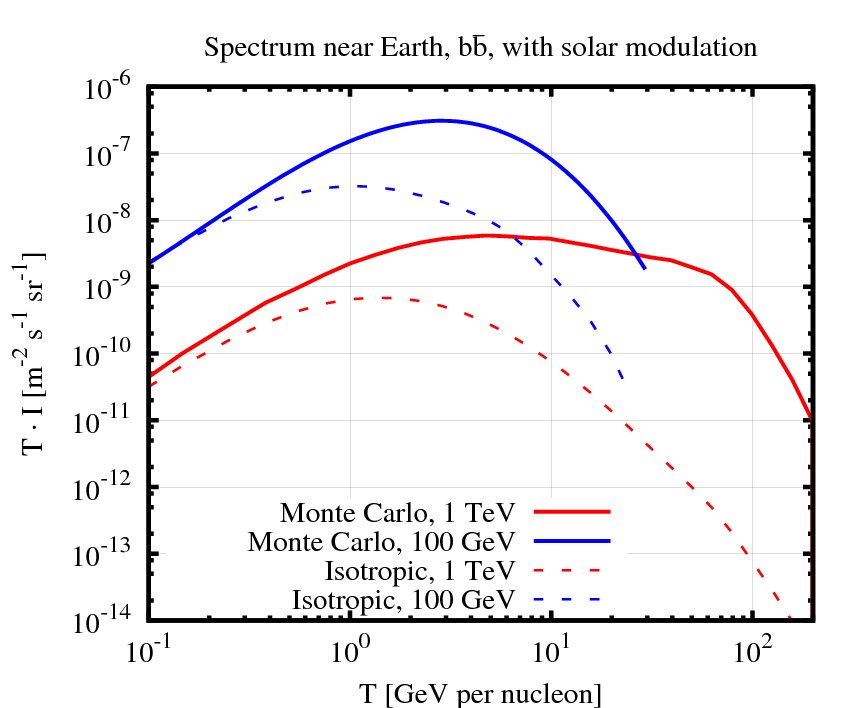}
\caption{Antideuteron spectra near Earth after propagation and Solar modulation. Calculations are done for dark matter masses of 1 TeV and 100 GeV, using the NFW density profile and the `med' propagation parameters. In both plots, we assumed a thermally averaged cross section of $\left< \sigma v \right> = 3\times 10^{-26} \rm{cm^3/s}$. Continuous lines show the result for the Monte Carlo approach, while dashed lines show the result from the isotropic approach.}
\label{fig:Final_spectra}
\end{figure*}

The resulting intensity of antideuterons using the 'med' propagation parameters is shown in 
Fig.~\ref{fig:Final_spectra}.
We see that there is a significant enhancement in the peaks of the spectra 
going from the isotropic to the correct Monte Carlo approach. This enhancement 
is most significant for the 1\,TeV $W^+W^-$ case, where the peak in the Monte
Carlo approach is two orders of magnitudes higher than in the isotropic 
approach. 

\begin{figure*}[t]
  \includegraphics[width=0.45\textwidth]{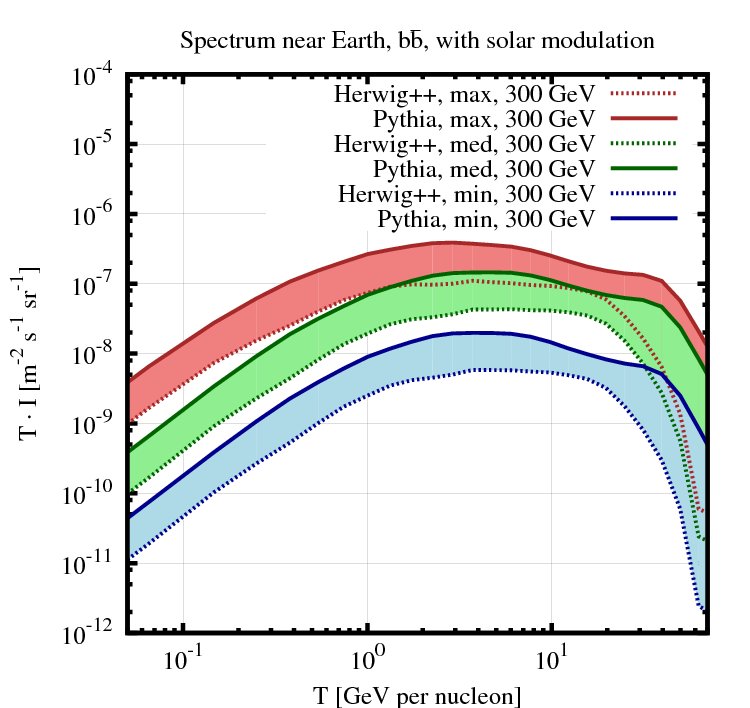}              
  \hfil
  \includegraphics[width=0.45\textwidth]{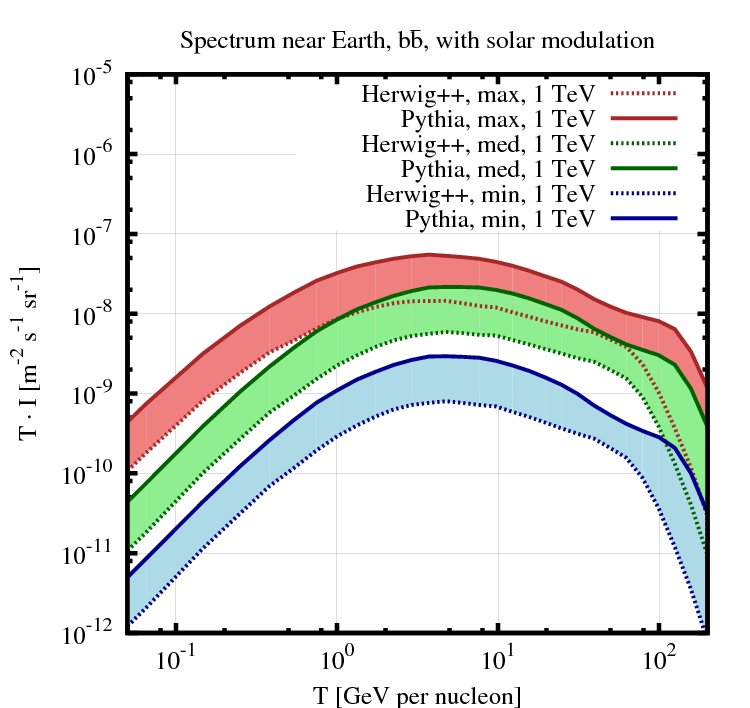}
  \includegraphics[width=0.45\textwidth]{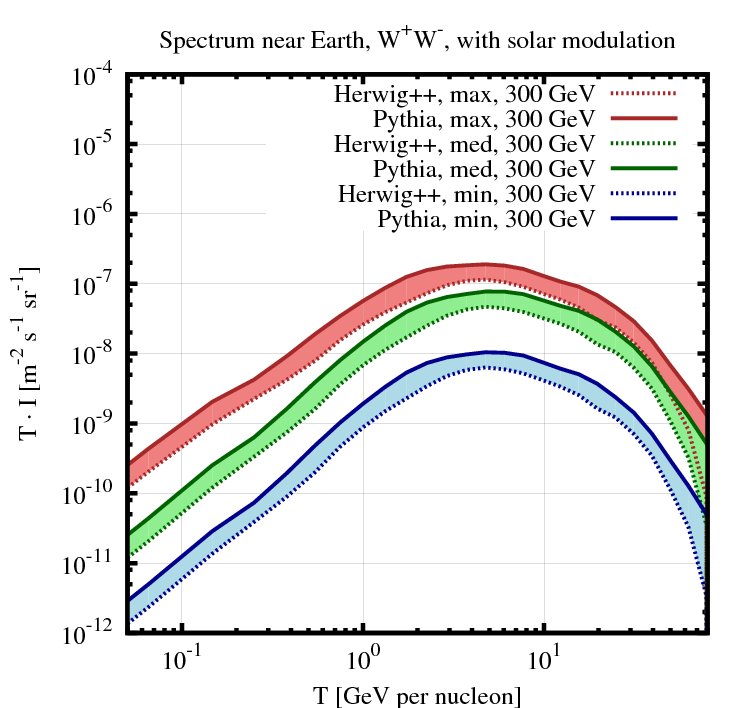}              
  \hfil
  \includegraphics[width=0.45\textwidth]{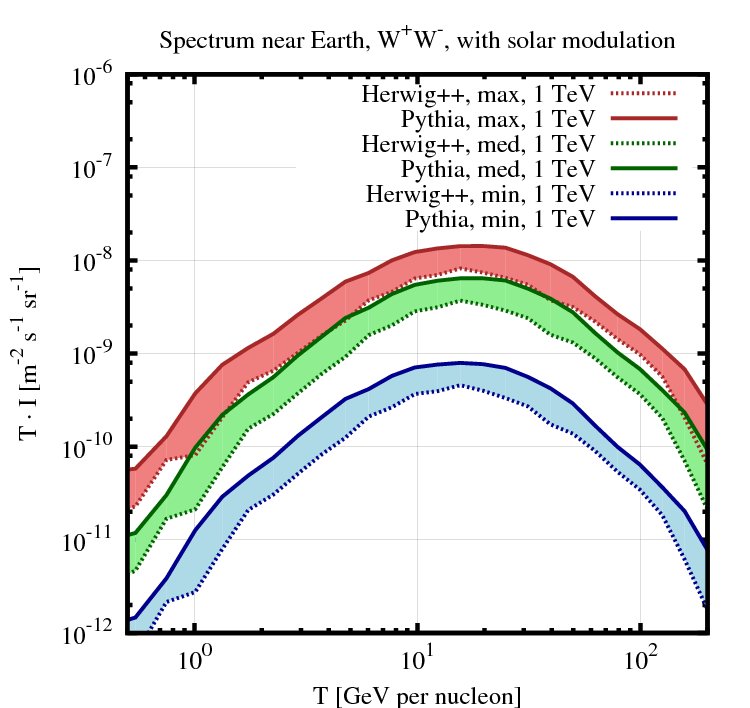}
\caption{Comparison of the antideuteron spectra from HERWIG++ and PYTHIA near Earth after propagation and Solar modulation. The plots show the results for dark matter masses of 1 TeV and 300 GeV, using the NFW density profile and varying propagation parameters. A thermally averaged cross section of $\left< \sigma v \right> = 3\times 10^{-26} \rm{cm^3/s}$. was assumed.}
\label{fig:Had_uncertainty}
\end{figure*}

In Fig.~\ref{fig:Had_uncertainty} we compare the intensities predicted by HERWIG++ and PYTHIA simulations, giving an estimate on the uncertainty in the hadronization.
From the figure, we see that the uncertainty 
in the propagation model leads to a spread in the final antideuteron 
spectrum of $\sim 1.5$ orders of magnitude, while the uncertainty in hadronization generally leads to an uncertainty of a factor 2--4, depending on the process in question.
For most energies, the uncertainty in the propagation clearly dominates over the uncertainty from hadronization. 
For the quark channel, however, the uncertainty in hadronization becomes competitive for energies corresponding to $x > 10^{-1}$.
Because of the shift in kinetic energy due to solar modulation, the sharp rise in uncertainty at low energies in the $W^+W^-$ channel is not visible in Fig.~\ref{fig:Had_uncertainty}, even with extended plot ranges.
We expect, however, that this uncertainty will become important for relevant ranges of the spectrum for DM masses in the TeV range.

%%%%%%%%%%%%%%%%%%%%%%%%%%%%%%%%%%%%%%%%%%%%%%%%%%%%%%%%%%%%%%%%%%%%%%%%%%%%%%
\section{Conclusions}

We calculated the antideuteron yield in dark matter annihilations in the 
coalescence model on an event-by-event basis and presented the resulting 
antideuteron fluxes for quark and gauge boson final states.
We showed that deuteron production is  very sensitive to the hadronization 
model employed. 
Comparing our results using the HERWIG++ Monte Carlo 
simulation to earlier results using PYTHIA, we found that the predicted
antideuteron yield varies by a factor $\sim3$ for $x\lsim 0.1$, $T\gsim 0.01$\,GeV the quark channel, and a factor $\sim2$ in the range $10^{-3}\lesssim x \lesssim 0.5$ in the gauge boson channel.
Outside these ranges, the uncertainty increases rapidly in both channels.
Varying the parameters of the cluster hadronization model employed in
HERWIG++  within the range suggested in Ref.~\cite{Richardson:2012bn} 
does not lead to significant additional differences in the antideuteron 
yield, taking into account that a recalibration of $p_0$ absorbs a pure
change in the  antideuteron multiplicity. 
Using the differences between HERWIG++ and PYTHIA as an estimate 
for the uncertainty introduced by the hadronization models, 
this results in an uncertainty of a factor of 2--4 on the observable intensity.
While this is generally subdominant to the uncertainty from the propagation model at around 1.5 orders of magnitude,
it is potentially dominant for the gauge boson channel at low energies for DM masses in the TeV range and above.
We also showed the importance of $n>2$ final states for
annihilations of DM particles with masses $\gsim 10$\,TeV.

%%%%%%%%%%%%%%%%%%%%%%%%%%%%%%%%%%%%%%%%%%%%%%%%%%%%%%%%%%%%%%%%%%%%%%%%%%%%%%
\acknowledgments

We are grateful to Sergey Ostapchenko and Are Raklev for  useful discussions,
and to Pat Scott for comments on the text.

%%%%%%%%%%%%%%%%%%%%%%%%%%%%%%%%%%%%%%%%%%%%%%%%%%%%%%%%%%%%%%%%%%%%%%%%%%%%%%


\begin{thebibliography}{99}

\bibitem{GKH}
K.~Griest and M.~Kamionkowski,
%``Unitarity Limits on the Mass and Radius of Dark Matter Particles,''
Phys.\ Rev.\ Lett.\  {\bf 64}, 615 (1990);
%%CITATION = PRLTA,64,615;%%
L.~Hui,
%``Unitarity Bounds and the Cuspy Halo Problem,''
Phys.\ Rev.\ Lett.\  {\bf 86}, 3467 (2001)
[astro-ph/0102349].
%%CITATION = PRLTA,86,3467;%%


%\cite{Donato:1999gy}
\bibitem{Donato:1999gy}
  F.~Donato, N.~Fornengo and P.~Salati,
  %``Anti-deuterons as a signature of supersymmetric dark matter,''
  Phys.\ Rev.\ D {\bf 62} (2000) 043003
  [hep-ph/9904481].
  %%CITATION = HEP-PH/9904481;%%


%\cite{Kadastik:2009ts}
\bibitem{Kadastik:2009ts}
  M.~Kadastik, M.~Raidal and A.~Strumia,
  %``Enhanced anti-deuteron Dark Matter signal and the implications of PAMELA,''
  Phys.\ Lett.\ B {\bf 683} (2010) 248
  [arXiv:0908.1578 [hep-ph]].
  %%CITATION = ARXIV:0908.1578;%%

\bibitem{master}
L.~A.~Dal, ``Antideuterons as Signature for Dark Matter'',
master thesis (Norwegian University of Science and Technology, 2011),
\url{http://urn.kb.se/resolve?urn=urn:nbn:no:ntnu:diva-12634}.

%\cite{Gieseke:2011na}
\bibitem{Gieseke:2011na} 
  S.~Gieseke {\it et al.}, 
%, D.~Grellscheid, K.~Hamilton, A.~Papaefstathiou, S.~Platzer, P.~Richardson, C.~A.~Rohr and P.~Ruzicka 
  %``Herwig++ 2.5 Release Note,''
  arXiv:1102.1672 [hep-ph].
  %%CITATION = ARXIV:1102.1672;%%

%\cite{Sjostrand:2007gs}
\bibitem{Sjostrand:2007gs} 
  T.~Sjostrand, S.~Mrenna and P.~Z.~Skands,
  %``A Brief Introduction to PYTHIA 8.1,''
  Comput.\ Phys.\ Commun.\  {\bf 178}, 852 (2008)
  [arXiv:0710.3820 [hep-ph]].
  %%CITATION = ARXIV:0710.3820;%%

\bibitem{old}
A.~Schwarzschild and C.~Zupancic,
  %``Production of Tritons, Deuterons, Nucleons, and Mesons by 30-GeV Protons on A-1, Be, and Fe Targets,''
  Phys.\ Rev.\  {\bf 129}, 854 (1963).
  %%CITATION = PHRVA,129,854;%%

\bibitem{Ka80}
J.~I.~Kapusta,
  %``Mechanisms for deuteron production in relativistic nuclear collisions,''
  Phys.\ Rev.\ C {\bf 21}, 1301 (1980).
  %%CITATION = PHRVA,C21,1301;%%

%\cite{Donato:2008yx}
\bibitem{Donato:2008yx}
  F.~Donato, N.~Fornengo and D.~Maurin,
  %``Antideuteron fluxes from dark matter annihilation in diffusion models,''
  Phys.\ Rev.\ D {\bf 78} (2008) 043506
  [arXiv:0803.2640 [hep-ph]].
  %%CITATION = ARXIV:0803.2640;%%

%\cite{Brauninger:2009pe}
\bibitem{Brauninger:2009pe}
  C.~B.~Brauninger and M.~Cirelli,
  %``Anti-deuterons from heavy Dark Matter,''
  Phys.\ Lett.\ B {\bf 678} (2009) 20
  [arXiv:0904.1165 [hep-ph]].
  %%CITATION = ARXIV:0904.1165;%%


%\cite{Schael:2006fd}
\bibitem{Schael:2006fd}
  S.~Schael {\it et al.}  [ALEPH Collaboration],
  %``Deuteron and anti-deuteron production in e+ e- collisions at the Z resonance,''
  Phys.\ Lett.\ B {\bf 639} (2006) 192
  [hep-ex/0604023].
  %%CITATION = HEP-EX/0604023;%%


\bibitem{pre}
 D.~Amati and G.~Veneziano, Phys.\ Lett.\ B 83, 87 (1979); Yu.~L.~Dokshitzer and S.~I.~Troian, Leningrad Nuclear Physics Institute preprint N922 (1984).

\bibitem{lund}
B.~Andersson,
{\em The Lund Model},
Cambridge Monographs on Particle Physics, Nuclear Physics and Cosmology
(Cambridge University Press 1994).

\bibitem{ellis}
R.~K.~Ellis and W.~J.~Stirling and B.~R.~Webber,
{\em QCD and Collider Physics}, 
Cambridge Monographs on Particle Physics, Nuclear Physics and Cosmology
(Cambridge University Press 2003).

\bibitem{Richardson:2012bn} 
  P.~Richardson and D.~Winn,
  %``Investigation of Monte Carlo Uncertainties on Higgs Boson searches using Jet Substructure,''
  arXiv:1207.0380 [hep-ph].
  %%CITATION = ARXIV:1207.0380;%%


%\cite{Cirelli:2010xx}
\bibitem{Cirelli:2010xx} 
  M.~Cirelli {\it et al.}, 
%G.~Corcella, A.~Hektor, G.~Hutsi, M.~Kadastik, P.~Panci, M.~Raidal and F.~Sala 
  %``PPPC 4 DM ID: A Poor Particle Physicist Cookbook for Dark Matter Indirect Detection,''
  JCAP {\bf 1103}, 051 (2011)
  [arXiv:1012.4515 [hep-ph]].
  %%CITATION = ARXIV:1012.4515;%%

%\cite{Alwall:2011uj}
\bibitem{mad} 
  J.~Alwall, M.~Herquet, F.~Maltoni, O.~Mattelaer and T.~Stelzer,
  %``MadGraph 5 : Going Beyond,''
  JHEP {\bf 1106}, 128 (2011)
  [arXiv:1106.0522 [hep-ph]].
  %%CITATION = ARXIV:1106.0522;%%


%\cite{Maurin:2001sj}
\bibitem{Maurin:2001sj}
  D.~Maurin, F.~Donato, R.~Taillet and P.~Salati,
  %``Cosmic rays below z=30 in a diffusion model: new constraints on propagation parameters,''
  Astrophys.\ J.\  {\bf 555} (2001) 585
  [astro-ph/0101231].
  %%CITATION = ASTRO-PH/0101231;%%

\bibitem{ewbrems}
This is a factor 100 lower than the value obtained considering only $Z$ emission 
from external fermion lines, cf.
V.~Berezinsky, M.~Kachelrie{\ss} and S.~Ostapchenko,
  %``Electroweak jet cascading in the decay of superheavy particles,''
  Phys.\ Rev.\ Lett.\  {\bf 89}, 171802 (2002)
  [hep-ph/0205218].
  %%CITATION = HEP-PH/0205218;%%


%\cite{Donato:2003xg}
\bibitem{Donato:2003xg}
  F.~Donato, N.~Fornengo, D.~Maurin and P.~Salati,
  %``Antiprotons in cosmic rays from neutralino annihilation,''
  Phys.\ Rev.\ D {\bf 69} (2004) 063501
  [astro-ph/0306207].
  %%CITATION = ASTRO-PH/0306207;%%

%\cite{Navarro:1995iw}
\bibitem{Navarro:1995iw}
  J.~F.~Navarro, C.~S.~Frenk and S.~D.~M.~White,
  %``The Structure of cold dark matter halos,''
  Astrophys.\ J.\  {\bf 462} (1996) 563
  [astro-ph/9508025].
  %%CITATION = ASTRO-PH/9508025;%%

\bibitem{Schopper1988}
   A.~Baldini, V.~Flaminio, W.~G.~Moorhead and D.~R.~O.~Morrison, 
   {\it Reaction 531~-~599}, H.~Schopper~(ed.), 
   SpringerMaterials - The Landolt-B\"{o}rnstein Database,
   DOI:~10.1007/10367917\_24

%\cite{Amsler:2008zzb}
\bibitem{Amsler:2008zzb}
  C.~Amsler {\it et al.}  [Particle Data Group Collaboration],
  %``Review of Particle Physics,''
  Phys.\ Lett.\ B {\bf 667} (2008) 1.
  %%CITATION = PHLTA,B667,1;%%


%\cite{Gleeson:1968zza}
\bibitem{Gleeson:1968zza}
  L.~J.~Gleeson and W.~I.~Axford,
  %``Solar Modulation of Galactic Cosmic Rays,''
  Astrophys.\ J.\  {\bf 154} (1968) 1011.
  %%CITATION = ASJOA,154,1011;%%

\end{thebibliography}
\end{document}